\documentstyle[aps,preprint]{revtex}
\def\beq{\begin {equation}}
\def\eeq{\end {equation}}
\def\beqa{\begin {eqnarray}}
\def\eeqa{\end {eqnarray}}
\def\pd{\partial}
\def\tpsi{\tilde{\psi}(\vec r, t)}
\def\psit{\psi(\vec r, t)}
\def\vr{\vec r}
\def\vrp{{\vec r_1}}
\def\a{{\sl a}}
\def\si{\sigma}
\def\phir{\phi(\vec r)}
\begin{document}

\title{ Planck Scale Physics of the Single Particle Schr\"{o}dinger Equation with Gravitational 
Self-Interaction}
\author{Vikram Soni}
\address{National Physical Laboratory, K.S. Krishnan Marg, New Delhi,
110016, India, vsoni@del3.vsnl.net.in}
\maketitle

\begin{abstract}
We consider the modification of a single particle Schr\"{o}dinger equation by
the inclusion of an additional gravitational self-potential term which follows
from the prescription that  the' mass-density'that enters this term is given by
$m |\psi (\vec {r},t)|^2$, where $\psi (\vec {r}, t)$ is the wavefunction and
$m$ is the mass of the particle.  This leads to a nonlinear equation, the '
Newton Schrodinger' equation, which has been found to possess stationary
self-bound solutions, whose energy can be determined exactly using an
asymptotic method.  We find that such a particle strongly violates
superposition and becomes a black hole as its mass approaches the Planck mass.
\end{abstract}

\section{Introduction}
Inspite of tremendous success of quantum mechanics  its interpretational
aspects continue to puzzle us, particularly in regard to measurements and
extrapolation to macroscopic levels.
The process of measurement involves a collapse of the wavefunction to 
one of the eigenstates of the operator being measured, and is
not describable within the framework of the normal unitary evolution
of the quantum mechanics. Further the quantum description when
extrapolated to macroscopic domain leads to paradoxical situations,
which arise when we superpose macroscopically distinguishable
quantum states. These two problems are related to each other, as the
measurement process at some stage or the other involves a macroscopic
apparatus and a result of the measurement by definition is a definite
number resulting from a "pointer position". This precludes states which
are superpositions of distinguishable pointer positions. The
pointer position is not predictable, and is found at one of the
possible positions whose
probability is dictated by its quantum wavefunction. The very nature of
our classical description is thus incompatible with quantum
superpositions and attendent probabilistic interpretation.

There have been suggestions that gravity has a possible role in the
fact that spatial superpositions of macroscopic objects are not seen.
Our paper, which was stimulated by some recent suggestions 
by Penrose, is also an attempt in this direction. 
Central to this set of proposals is the idea that the wavefunctions 
that involve superpositions of spatially separated wavepackets should 
also have superpositions of the gravitational fields associated with 
the distinct mass distributions of the wavepackets. As has been lucidly
explained by Penrose, there are basic difficulties in combining the
covariance principle of general relativity and superposition principle
of quantum mechanics. The difficulty is best seen in studying the 
superposition of states that are spatially apart.
The gravitational field states involved in superposition require
different space-times among which a point-wise identification is not
possible.  Under these conditions, it is impossible to
define a unique time-translation operator and the very concept of stationary
state. According to Penrose this leads to an uncertainity of energy,
which makes such spatially superposed states inherently unstable.

At another level, heuristically, if we assume that a particle may be
localised at most in a size given by its Compton wavelength, then for
a particle whose mass becomes of the order of the Planck mass or more and
whose size is taken to be of the order of its Compton wavelength, the particle
could be a black hole. This suggests that there is perhaps a problem with
the QM of such particles.

One way to model these effects may be to include a non linear mass dependent
gravitational potential energy term in the standard Schroedinger equation,
by postulating a 'mass density' given by
 $m |\psi (\vec {r},t)|^2$, where $\psi (\vec {r}, t)$ is the wavefunction
 and $m$ is the mass of the particle. This term will automatically violate
 superposition.(  we cannot give any a priori reason for gravity to
 violate superposition and generate such a term in the Schroedinger equation).

 Such a nonlinear modified equation, called the Newton Schroedinger equation
 has a long history \cite{r1,r2,r3,r4,r5,r6,r7,r8,r9,r10,r10a,r11,r12,r13} in the
  context of works that
look at the influence of gravity on Superposition in QM. This equation has
been used in several many body contexts, such as Hartree Fock calculations in
plasmas as well as in astrophysics \cite{r14,r15,r16}
. (It is also known as Choquard's equation \cite{r17} and the Schroedinger-Poisson
equation \cite{r18} . The existence of stationary solutions has been
known  (see for example \cite{r17}) for sometime) 
In a preceding paper \cite{r19}, we looked at that the 
  energy eigenstates of the equation and using an asymptotoic method could
  find the exact eigenvalue \cite{r19}.We review this in Sec. I, II and III.    

  In Sec. IV (A and B) we take up the question of the spatial superposition of
  such localized stationary solutions and compute the energy difference
  between the superposed and stationary states , which may be identified with the
  breakdown of superposition. We show that this is related to an energy
  uncertainty calculated by Penrose using the the difference in the free
  fall accelarations corresponding to the the mass distributions of the two
  distinct spatial components of the superposition.This energy difference
  is used in analogy to Penrose's \cite{r13} calculation to get a time for
  state reduction for an arbitrary mass particle.

 In Sec. IV (C) we first observe that our NS equation is valid only up to a
 limiting mass and not beyond. This follows from the fact that the
 bound state energy of the stationary state goes as $ m^5$  and thus
 overtakes the rest mass at a certain value of the mass causing our
 particle to be unstable.

 For masses just below the limiting mass, using our  exponentially localised
 wave function, we can calculate the expectation value of the radius of our
 bound state particle of mass ,m. On substituting this radius in the
 expression for the horizon parameter  we find that the particle becomes a
 black hole somewhat below the limiting mass.We also find that superposition
 is  strongly broken for such values of the mass.

\section{Gravitational self-interaction}
The implementation of the above set of semiclassical and
nonrelativistic approximations amounts to saying that the particle 
experiences a self-gravitational potential, which arises
from the gravitational potential energy due to the mass density
given by $m|\psi(\vec r, t)|^2$, where $\psi(\vec r, t)$ and $m$ are the
wavefunction and mass of the particle respectively. Incorporating this
interaction in the Schr\"{o}dinger equation, one obtains the following
equation.
\beq
i\hbar {\pd \psit \over \pd t} = -{{\hbar}^2 \over 2m} {\nabla}^2
\psit + V(\vec r) \psit + m V_G(\vec r) \psit
\eeq
where $V(\vec r)$ is the external potential acting on the particle and
$V_G(\vec r)$ is the gravitational self-potential arising due to mass
density obtained from the wavefunction of the particle itself. Thus
$V_G(\vec r)$ is given by,
\beq
V_G(\vr) = -G \int {m|\psi(\vrp,t)|^2 \over |\vr - \vrp|}
d^3 r_1
\eeq
and equivalently,
\beq
{\nabla}^2 V_G(\vr) = 4 \pi Gm|\psit|^2
\eeq
 In order to get an idea of the magnitude of self-coupling,
let us consider the external potential , V, to be the Coulomb potential,
$ V = -e^2/r $.
We can write the
equation in a dimensionless form by introducing a free  length  $\a$. In terms
of this, the time is measured in units of $\tau = 2m{\a}^2/\hbar$, and
the energy is measured in units of $\epsilon = \hbar/\tau = {\hbar}^2/
2m{\a}^2$. Let, 
\beqa
t = \tau \tilde{t};  \;\; r = \a \tilde{r} \\
E = \epsilon \tilde{E};  \;\; \psi = \tilde{\psi}/{\a}^{3 \over 2}
\eeqa
In these units, our equations take the form,
\beq
i{\pd \tpsi \over \pd t} = - {\nabla}^2 \tpsi -
{-2m{e^2} \a \over {\hbar}^2} {1 \over \tilde{r}} \tpsi +
\tilde{V}_G(\vec r) \tpsi
\eeq
and
\beq
{\nabla}^2 \tilde{V}_G(\vr) = 4 \pi C |\tpsi|^2
\eeq
where $C$ is a dimensionless coupling constant given by,
\beq
C = {2Gm^{3}\a \over {\hbar}^2}
\eeq

By choosing the length scale  'a'to set the coefficient of the Coulomb term
to be unity, that is  $ a = {\hbar}^2/{ 2m{e^2} } $ we can immediately read
off the value of the dimensionless parameter , $ C = {G{m^2} }/{e^2} =
 {m^2/m_{pl}^2 \over {\alpha}} $, where $ \alpha $ is the fine structure
 constant. This clearly shows that the additional gravitational term
 we have added is relatively negligible compared to the coulomb interaction
 unless masses get rather large ; m of the order of $ m_{pl} \sqrt{\alpha} $ .

Before we examine the solutions of this nonlinear equation, it is
useful to note the two conservation equations for particle number (or mass)
 and
energy.These have been already noted in \cite{r5}. The first equation that is easily established is for particle density,
$\rho (\vr,t) = {\psi}^*(\vr,t) \psit$. It obeys the usual continuity
equation given by,
\beq
{\pd \rho \over \pd t} + {\vec \nabla}.{\vec J}_p = 0 
\eeq
In this and the following equations, we shall be working in units
defined above, without explicitly putting tilde over dimensionless
quantities. Next we consider the energy functional $E(\psit)$ ,
 \beq
E = \int [|\vec \nabla \psit |^2 + V(\vr)|\psit|^2 + {1 \over 2}
V_G(\vr) |\psit|^2 ]
\eeq
The time-dependence of $E(\psit)$ satisfies,
\beq
{\pd E \over \pd t} + \int d^3r {\vec \nabla}.{\vec J}_E = 0
\eeq
where $J_E$ is given in \cite{r19}.
%\beq
%\vec J}_E = i[({\nabla}^2 \tpsi^*) {\vec \nabla} \tpsi - {\vec \nabla} \tpsi^*
%({\nabla}^2 \tpsi)] + i[({\vec \nabla} \tpsi^*) \tpsi - ({\vec \nabla}
%\tpsi) \tpsi^* ] [ V(\vr) + V_G(\vr)]
%\eeq
Note that energy density is quartic in $\psi $, and thus can not be
expressed as an expectation value of an operator.
We should also point out by calculating the expectation value of the
momentum operator we find the consistency that our gravitational self
potential does not contribute and therefore not give rise to a force
or any spurious motion \cite{r5,r19}.
\section{Stationary Solutions}
Solutions to this equation have been considered from a variational point of
view in \cite{r5} and  numerically in \cite {r20} and more recently
in \cite{r21}.
We now show that S-N equation admits stationary solutions ,
as found in \cite{r19} , of the form,
\beq
\psit = e^{-iEt} \phir
\eeq
The reason is that for solutions of this form the self-potential $V_G$ becomes
time-independent. $\phir$ obeys the equation,
\beq
E \phir = - {\nabla}^2 \phir + V(\vec r) \phir - C \int
{{|\phi(\vrp)|}^2 \over |\vr - \vrp|}d^3  r_1 \phir
\eeq

We now  present an analysis of the equation in the absence of an
external potential. 

The time independent equation also admits a variational interpretation,
and can be regarded as an extremum of the functional,
\beq
H[\phir] = {1 \over 2}\int d^3r |{\vec \nabla }\phir |^2 - {C \over 2}
\int \int {{|\phir|}^2{|\phi(\vrp)|}^2 \over |\vr - \vrp|}
d^3 r_1 d^3r
\eeq

with the normalisation constraint  $\int {{|\phir|}^2}d^3 r = 1 $.

It is useful to record that with this constraint the energy functional above
,under the scaling of  $  r \rightarrow  br $, has a scaling behaviour that
yields a minimum under the variation of b , indicating that the solution 
is stable with respect to scaling .

 We  analyse one class of
solutions of the free particle equation by doing an asymptotic
analysis \cite{r19}. This procedure allows us to obtain
eigenvalues ${E}$ for these solutions exactly, but the wave function is
known only in the region of large $r$. The key observation is that if
$\phi$ is taken to be an exponentially localised function about an
arbitrary point, it generates an asymptotic  potential $V_G$ which
is a monopole potential $C/r$ with corrections that decay exponentially
with $r$, where $r$ measure the distance from the point around which
$\phi$ is localized.

	These remarks are best illustrated by taking $\phir$ to be the
ground state hydrogenic wave function $(1/{\pi \si^3})^{1 \over 2}
e^{- r/\si}$. The corresponding self-potential is,
\beq
V_G(r) = -C {\big [} {1 \over r} - e^{-2r/\si} {\big (}{1 \over r} +
{1 \over \si} {\big )]}
\eeq
Substituting the solution in Eq.(15) and keeping terms to order
$e^{-r/\si}$, one finds,
\beq
E e^{-r/\si} = ({2 \over \si r} - {1 \over {\si}^2})e^{-r/\si} - {C \over
r}e^{-r/\si}
\eeq
which yields value of $E$ and $\si$ to be,
\beq
\si = {2 \over C}  \; \; and \; \;  E = -{1 \over {\si}^2} = - C^2/4 
\eeq

The asymptotic wave function and the potential gets exponential corrections
that are exactly calculable while the eigenvalue is not changed ( see \cite{r19}
 for details ).
(Note that here 
 'r' or $ \si $ in  units of $\a$
, the time in units of $\tau = 2m{\a}^2/\hbar$ and
the energy  in units of $\epsilon = \hbar/\tau = {\hbar}^2/
2m{\a}^2$.)   

\section{Physical Implications}
In this section we consider some physical implications of the results
obtained above. The above results show that the gravitational
self-potential leads to a self bound state of linear extent of order
${1 \over C}$. Again for a microscopic particle like the electron this
means a localisation over enormous distances of order $10^{43} A$, with
a binding energy of $10^{-87}$ Ryd. Thus on scales much smaller than 
$\si$, we do not expect the gravitational potential to affect things, 
and the usual quantum mechanics should apply.

\subsection{SUPERPOSITION }

In  the context of  the stationary states of the NS equation we can examine
the existence of spatial
superpositions or cat states in somewhat greater detail. Note that
the centre of the stationary localised states discussed above can be chosen
arbitrarily as the free particle equation is translationally invariant. More
specifically, we calculate the energy of a cat-state consisting of
two self-localised humps with separation much larger than $\si$.  The
wavefunction of such a state is,
\beq
\phi _c(r) = [\alpha \phi_s(r) + \beta \phi_s(|\vr -\vec b|)]
\eeq
There is a subtle difference between the localised states of our Newton
Schroedinger equation and those of ,for example, an Hydrogen atom.
In our case there is no separate centre of mass (CM) coordinate , whereas
in the case of the H atom there is a localised stationary state for the
relative coordinate and CM wavefunction is free to be in a plane wave or
localised state.

 It would then be appropriate to identify our stationary wave function
  with a coordinate/
 localised state of the CM of the Hatom. Since the H atom has very
 large mass compared to mass of the electron, localising the atom costs
 far less energy than localizing the electron. If we then neglect the
 former , we can effectively ( i.e. approximately) use the relative WF
 by itself to describe the localised atom.This we do.

First consider the case of the usual Quantum Mechanics with linear
 operators which is based on the Superposition principle ( that is in
 the absence of the non linear gravitational interaction term). In this
 case if we identify $ \phi_s(r)$ as the usual lowest s wave state of the
 Hydrogen atom , that is a stationary state with energy eigenvalue $E_0$
 centred at the origin, then  $ \phi_s(|\vr -\vec b|)$ is the same state
 centred at   $ \vec b $. If the separation $|b| >> \si $,
the scale of the wavefunction , the superposed state above has the same
energy eigenvalue.

Now consider the case in the presence of the non linear gravitational
interaction term, and let  $\phi_s$ denote the  stationary state
solution derived above. When $b>>\si$ the function is normalized with
${\alpha}^2 + {\beta}^2 = 1$. Under this condition the energy of the
superposed state is  not the same any more.This is because
i) the superposed state is no longer a solution of the non linear NS
equation and ii) the gravitational potential energy between the lumps
comes into play. In fact for  asymptotic
separation between the lumps it is always higher,given by:
\beq
E_c = T - ({\alpha}^4 + {\beta}^4)K_{11} - 2{\alpha}^2{\beta}^2 K_{12}
\eeq
where $T$ and $K_{11} = V_G  $ are the kinetic and potential energies of the
single stationary state , and $K_{12}$ is the gravitational potential energy
 between
the humps. Since $K_{12}$ is of order $1/b$, it is much smaller than
$K_{11}$ (for asymptotic separations)  and can be neglected.
 It is then easily seen that 
energy ofthe superposed state is higher than the energy
of the single stationary state and in this asymptotic case it is clear that
this
originates not from the gravitational potential energy between the humps
,but from the normalization of the superposed state and the non
 linearity ( in its dependance on the wave function) of the gravitational 
interaction

We specifically choose $ \alpha $= $\beta $ = $1/\sqrt{2}$ to conform to
the example considered by Penrose  \cite{r13}. In this case for asymptotic
separation we get a factor of 2 in the numerator ,corresponding to two lumps
and a factor of $1/4$ from the normalization , giving a potential energy
 equal to $ K_{11}/2 $ (neglecting the $ K_{12} $
term as we consider asymptotic separations ) for the superposed state,
 which is exactly half of that for the single lump stationary state.

 It is clear from the solution that the
 $ |E|  =| V_G/2| = |K_{11}/2| $ ; this is also a simple consequence of the
  virial
  theorem. The energy difference, $ \Delta V $ , between the superposed state
 and the stationary state is then given by , $ \Delta V = |K_{11}/2| = |E| $.
 This is a measure of sorts of the breakdown of Superposition in the NS
 equation, as it is zero for the case of linear Quantum Mechanics.

\subsection{The Penrose conjecture on the time of state reduction of
 Superposed states}

  Penrose \cite{r13} has argued that such superposed states are unstable due to energy
uncertainty arising from  the  mismatch of the two space times
due to the two lumps respectively that make up the superposition.
 
More specifically a typical energy scale is obtained by Penrose by
integrating the modulus squared of the difference of the free fall
accelarations due to each of the two localised lumps that make the coordinate
superposition of the stationary states. This energy turns out to be the
gravtational  energy of the difference of the mass distributions
due to the two lumps.Penrose chooses to term this the gravitational self
energy (GSE) of the difference.We will use the same acronym - GSE.

A typical time of superposed state reduction is then obtained by
dividing $\hbar$ by this energy in analogy with the decay of an
unstable particle.
These observations are general , for any given mass distributions -not
specific to those considered above for the NS particle.

 We shall now show that
there is a close connection between the GSE computed by Penrose \cite{r13}
(and others) and the energy difference between the superposed state
and the stationary state , $ \Delta V $ , above.  Assume that the
' mass density' for  a particle governed by the NS equation is given by ,
 $\rho_s(\vec r)$ = $m|\phi_s( r)|^2$ .

On the other hand the normalized superposed wavefunction is 
\beq
  \phi_c(r) = {1\over \sqrt{2}} [  {\phi^1}_s(r) +  {\phi^2}_s(r)]
\eeq

where
 $ \sqrt{{\rho^1}_s(\vec r)/m} = {\phi^1}_s(r)  = \phi_s(|\vr|)$

and $ \sqrt{{\rho^2}_s(\vec r)/m}={\phi^2}_s(r)=  \phi_s(|\vr -\vec b|)$

 The mass density for the superposed state is 
$\rho_c(\vec r)$ = $m|\phi_c( r)|^2$

We can then write down the gravitational  energy of the 
superposed state
\beq
V_c = - {C\over 2}
\int \int {{|\phi_c(r)|}^2{|\phi_c(r\prime)|}^2 \over |\vr - \vr\prime|}
d^3 r d^3r\prime
\eeq

In the asymptotic limit , that is when $ |b| >> \si $ , there are terms in
the expression above that are exponentially suppressed by the argument
 $ -|b|r  $ and can be dropped.In the neglect of these terms and with
some routine juggling we can get the energy difference between 
the superposed state $\phi _c(r)$ and the stationary state $\phi_s( r)$,
$ \Delta V = V_c - V_G $
\beq
\Delta V = {1\over 4} {G\over 2}
\int \int {(({\rho^1}(r))({\rho^1}(r'))+({\rho^2}(r))({\rho^2}(r'))- 
({\rho^2}(r))({\rho^1}(r'))-({\rho^1}(r))({\rho^2}(r')) \over |\vr - \vr'|}
d^3 r d^3r'
\eeq
which is proportional to the GSE $\Delta$ that
appears in \cite{r13}        

\beq
\Delta = {4\pi} {G}
\int \int {(({\rho^1}(r))({\rho^1}(r'))+({\rho^2}(r))({\rho^2}(r'))- 
({\rho^2}(r))({\rho^1}(r'))-({\rho^1}(r))({\rho^2}(r')) \over |\vr - \vr'|}
d^3 r d^3r'
\eeq

This establishes the connection between the  energy difference,
$\Delta V$ ( which is the the difference of the gravitational  energies
 ( GSE )  between the the superposed (asymptotic) and stationary state )
that is a measure of the breakdown of the superposition principle for us
and the energy uncertainty,$ \Delta $,  (which is propotional to the GSE
 of the density differences of the two humps ) associated
 with superposition in \cite{r13}.

    There are problems with the above identification.
    We should note here that it is obvious that the mass density
    expression used in constructing the gravtational term for the NS
    equation is not in any sense mean that the electron mass is distributed
    according to its wavefunction it can have only the usual QM probalilistic
    interpretation.However as the mass of the particle is increased its
    behaviour becomes more and more classical and such an interpretation may
    become plausible.( Anyhow, we expect that if the asymptotic separation
  parameter b is of the order of laboratory scales of centimeters ,and we
 constrain the stationary wavefunction to be spatially smaller ,say $ 10^{-5}
  cm $. This implies that realistically we are dealing with particles of a
  mass in  excess of  $ 10^{12} Gev $.)

 For a particle of mass ,m ,
governed by the NS equation we importantly find that  ,$ \Delta V = |E| $,
goes as $ m^5 $

This is indeed different from the classical expression for the GSE of
a constant density mass distribution which goes as  ,$ m^2/R $ ,which in turn
goes as $m^{5/3} $ , for composite objects
like droplets considered in \cite{r22}

(We observe that these expressions for , $ \Delta V $  and $ \Delta $
are proportional to the GSE for the single statioary state  because, for
asymptotic separation, the gravitational potential energy between the lumps
can be neglected. )

On assuming Penrose's conjecture  that this is the energy uncertainty
associated with an unstable particle,
we can now explicitly find the time of state reduction for the
superposed state of a NS particle as a function of m . This is

       $ T_{SP}  =  \hbar/{\Delta V}  = \hbar/{|E|} = 2{\hbar}^3/{G^2{m}^5} $

For specificity this is

 i)$ 10^{70}$ sec, for an electron

 ii) $10^{-40}$ sec. for a particle of a mass that is less 
 than  the limiting mass by a factor of 5 -which puts it it safely in a
non relativistic ,stable and non black hole regime ( see following section ).

For comparision we can estimate the time of state reduction for a
droplet of constant density ,when the GSE given in \cite{r22} goes as
, $ m^{5/3}$.

      $ T_{SP}  =  \hbar/{\Delta } = \hbar/{G{m}^2/R} = {\hbar}/{G{m}^{5/3}} $

 This is   $10^{-10}$ sec. for a particle of a mass that is
 less than the limiting mass  by a factor of 5.

\subsection{ LIMITING MASS OF A PARTICLE }

It must be noted that the energy eigenvalue ( which is negative for the bound
state ) goes as  $ m^5 $ . 
This results in an instability as the sum of the rest mass energy and
 the attractive binding energy becomes negative. At this point our
 Shroedinger Newton description with only the gravitational interaction
 breaks down.
The value of this limiting mass is obtained on putting the ratio of the
 modulus of the energy eigenvalue and the rest mass , $ \Gamma $ = 1 
(Note  $ \Gamma $  must be small compared to unity for our analysis to work.)
\beq
m_u = ({2{\hbar}^2 c^2 \over G^2})^{1\over 4}
\eeq
 This is effectively the Planck mass (apart from a constant factor of O(1) )
Perhaps this is not so surprising for a theory which has only gravitational
interactions.

    We would like to consider more of the physics as we approach this limiting
 mass. To this end we determine some relevant parameters for particles of
 mass that approaches the limiting mass and compare them to their counterparts
 for the electron.
 We note that our theory has only one dimensionless
parameter, $\delta =  (m^2G/\hbar c) $. All dimensiomless quantities are
then expressed in terms of, $\delta $.

i) The parameter $ \Gamma =|E|/{m{c}^2}$ above goes as 
$ {1\over 2} (m^2G/\hbar c)^2 ={1\over 2}{\delta}^2 $. It is

 $ 1$  for the limiting mass ,

 $ 10^{-90}$ for the electron 
                                                                                                
 Recall that in our theory the breakdown of Superposition is linked to the
 energy , $\Delta V $ which  is further
 equal to , E , the energy eigenvalue of the bound state. Dividing by
 the rest mass energy yields, $ \Gamma =|E|/{m{c}^2}$ , which then is
 also a measure of sorts of  the breakdown  of superposition .

ii) The size parameter R

The typical size for a particle,
R ,may be evaluated by calculating the expectatation value of , 'r' ,
for the wave function $(1/{\alpha}^3 \pi)^{1 \over 2}
e^{- r/\alpha }$,where, 

$ \alpha = a \si  = a2/C = \hbar^2/{G {m}^3}$

Thus  ,  $ R =   { 3 \alpha \over 2}  $

iii) The Horizon parameter ,  $ 2Gm/{R{c}^2}$ = ${ 4 \over 3}(Gm^2/\hbar c)^2
 $
 =$ {4 \over 3} {\delta^2} $

where we have substituted for R above . We should further notice that
this parameter goes as  $m^4$ or the square of   $  \delta $.

This parameter decides if a mass distribution is a black hole or not.
Specifically, if it is greater than  1 we have a black hole.It is

 ${8 \over 3} $  for the limiting mass  
     
 $10^{-90}$  for an electron          

Clearly, at the limiting mass the particle is a black hole !

This analysis shows that in the presence of a gravitational modification 
of the Schroedinger equation as given above we find not only a stability 
problem with
masses as we approach the limiting mass but that such a particle would be
a black hole. Note however that a reduction of the mass by only a factor of
5 of the limiting mass gives a very sensible non relativistic description and
brings down the Horizon parameter by a factor of 125 taking it safely away
 from a black hole.

We have studied a certain cojectural non linear modification of the
non relativistic Schrodinger equation due to gravity - the Newton-
Schrodinger equation to look at the quantum mechanics of such a particle.
The implication is that in this description as we approach the Planck
mass not only do we strongly violate superposition but the particle
becomes a black hole, thereby putting a limit on the mass of
elementary particles.

{\bf Acknowledgements}

First I thank  Deepak Kumar with whom  the earlier part of this work was done.
  I am grateful to V.P.Nair, Vikram Vyas, G.Ghirardi, D.Diakonov and
H.Hansson for discussions and to E.R. Arriola for bringing to our attention
some earlier work done on similar equations.

\end{document}